\documentstyle[12pt]{article}
\begin{document}
\pagestyle{plain}
\hsize = 6.5 in 				
\vsize = 8.5 in		   
\hoffset = -0.6 in
\voffset = -0.5 in
\baselineskip = 0.22 in
\def \SN{\vskip 0.3 true cm \noindent}

\noindent
\today

\vskip 4.0cm \noindent
{\large{\bf An Analysis of the Thermodynamics of Hydrophobic
Solvation Based on Scaled Particle Theory}}

\vskip 0.5cm \noindent
Hong Qian

\vskip 0.5cm \noindent
Department of Applied Mathematics and Bioengineering
\vskip 0.0 true cm \noindent
University of Washington, Seattle, WA 98195

\vskip 1.5cm \noindent
{\large {\bf Synopsis}}
\vskip 0.5cm \noindent

	A comprehensive, semi-quantitative model for the thermodynamics 
of hydrophobic solvation is presented.  The model is based on a very
simple premise suggested by the scaled particle theory and treats both 
solute and solvent molecules as hard spheres.  A connection 
between the peculiarly large heat-capacity change for hydrophobic 
solvation and the large temperature dependence of the thermal 
expansivity of water is found.  Analysis reveals a possible physical 
origin for the converging behavior of solvation entropies for a 
series of homologous hydrophobic compounds.  The model suggests that the 
low solubility and the large heat-capacity change of hydrophobic solvation 
stem from two distinct aspects of water molecules: the static geometry 
of the molecule and the dynamic hydrogen bonding network, respectively.

\vskip 1.5cm \noindent
Running title: Hydrophobic Solvation Thermodynamics

\pagebreak

\vskip 1.0cm \noindent
{\large{\bf Introduction}}
\vskip 0.5cm \noindent

	The scaled particle theory (SPT), a statistical mechanical model 
for liquids and solutions made of hard spherical molecules, is a powerful 
conceptual framework and a useful computational tool for molecular 
solvation.$^{1-4}\ $  Even though the SPT was originally developed
only for hard spherical molecules,$^5\ $ recent studies have shown that 
it also provides insight into the solvation of hydrophobic solutes in 
associative liquid like water.  There are several reasons for this wide
applicability of SPT:

\SN 1) The SPT is not a statistical mechanical theory based on the first 
principle.  Rather, an independently obtained solvent density has to 
be provided in addition to temperature, pressure, and molecular 
parameters.  This is in contrast to a genuine statistical mechanical 
theory of liquid, in which density is derived from temperature, pressure, 
and molecular parameters.  Some essential properties of an associative 
liquid are contained in the density data.$^{5,6}\ $
 
\SN 2) Recent studies have shown that the SPT provides accurate numerical 
calculations for solvation free energy of hydrocarbons in water, but 
does so with less success for calculating the entropy and enthalpy.$^{2,6}\ $
This result is consistent with the concept of entropy-enthalpy compensation,
which occurs with the solvent reorganization when introducing a solute 
molecule into a solvent.$^{6-9}\ $

\SN 3) To apply the SPT to an associative liquid like water, one practical 
difficulty is how to determine the hard sphere radii for water
and the solute molecules.$^2\ $  While these difficulties affect 
quantitative numerical calculation, it does not alter the qualitative 
physical insight one obtains from the SPT.  In this paper, our primary
interest is in the qualitative relationship between different 
physical quantities, the numerical values are not essential for our 
analysis.

\vskip 0.3 cm \noindent

	Based on these arguments, we try to use SPT as a semi-quantitative
model for understanding various experimental results which are 
essential to the solvation of small organic molecules in aqueous 
solutions.  We show that much of the characteristics of hydrophobic
solvation can be understood in terms of this over-simplified 
model.  Our approach is very much in the same spirit of the earlier work 
of Grunwald.$^8\ $  Treating the hydrophobic solutes as inert hard spheres, 
of course, completely neglects the solute-solvent interaction.$^{3,6}\ $  
This should be kept in mind when applying the result of the present analysis 
to the solvation of hydrocarbons.  On the other hand, the dissolution of
inert gases like xenon exhibits all the characteristics of hydrophobic 
solvation.  Hence, neglecting solute-solvent interaction is justified 
in our present study.

\vskip 1.0cm \noindent
{\large{\bf Some Basic Facts}}
\vskip 0.5cm \noindent

	The essential features in the data for dissolution of hydrophobic
compounds in aqueous solutions are summarized as follows: 

\SN 1) Dissolution entropy changes, $\Delta S(T)$'s, associated with 
transferring a series of hydrophobic solutes 
from solid $\rightarrow$ water, liquid $\rightarrow$ water, and 
gas $\rightarrow$ water form three distinct, respective groups.
Within each group, $\Delta S(T)$'s for different solute species converge 
to a common $\Delta S^*$ at a temperature $T^*_s$.  While 
$\Delta S^*$'s are different for the three groups, $T^*_s$'s for all 
three groups are approximately equal to $110^{\circ}$C.$^{10}\ $
The two groups of data on the transfer from liquid $\rightarrow$ water 
and gas $\rightarrow$ water are in agreement with the Trouton's rule.

	In order to compare free energies calculated from SPT with 
experimentally obtained solubility, one has to calculate the solvation free 
energy from a measurement based on concentration scale in molarity
rather than in mole fraction.$^{5,11,12}\ $  The numerical difference between 
the two calculations is $-RTln(v_1/v_2)$ where $v_1$ and $v_2$ are the molar 
volumes of the solvent systems before and after the transfer, respectively.  
For the gas phase, $v = k_BT/p$ where $p$ is the gas pressure (= 1 atm); 
for water, $v$ =18.08 $cm^3/mol$.$^6\ $  This changing of concentration 
scale increases the free energy values given by Murphy et al.$^{10}\ $ 
by about 60 J per degree per mole$^{13}\ $ (e.g., shift the lines in their 
Figure 1 upward).

\SN 2) The heat-capacity changes, $\Delta C_p$'s, for all the dissolution 
reactions are approximately temperature independent.$^{14}\ $

\SN 3) There also appears to have a convergence temperature for enthalpy 
changes
in dissolution of organic compounds from gas $\rightarrow$ water (see 
Appendix for more discussion).  The temperature is not known, but the 
corresponding $\Delta H^*$ is about a few KJ/mol.

\vskip 0.3 cm \noindent

	There have been many different ways of presenting experimental 
data on the thermodynamics of dissolution.  Here we have summarized
several key results from the literature which are useful in organizing 
and relating various data representations.  It is important to note
that many presentations are in fact concerning same experimental
measurements.  They are related by the basic thermodynamic formulae,
as shown in the following theorems.

\vskip 0.3 cm
\SN {\it Baldwin's Theorem}
	
	Based on the observations of Sturtevant$^{15}$ and 
Privalov$^{14}$, R.L. Baldwin gave the following theorem in
1986.$^{16}\ $ For a series of homologous reactions, designated by
$i$, with temperature-independent $\Delta C_{p_i}$, the following 
two statements are equivalent:
\SN 1) At one temperature $T^{\dag}$, there are unique constants 
$a$ and $b$ such that:
\begin{equation}
             \Delta S_i = a + b\Delta C_{p_i} 
\end{equation} 
2) there is a unique temperature $T^*_s$ at which
$\Delta S(T^*_s)$ is unique.  That is, $T^*_s$ and $\Delta S_i(T^*_s)$ 
are independent of $i$.  Hence: 
\begin{equation}
      \Delta S_i(T) = \Delta S(T^*_s) + \Delta C_{p_i} ln(T/T^*_s)  
\end{equation} 
Compare Eq. (1) with (2), we have $a$ = $\Delta S(T^*_s)$ and 
$b$ = $ln(T^{\dag}/T^*_s)$.  From now on, we will follow Lee$^{13}\ $
and refer to the $\Delta S_i$ versus $\Delta C_{p_i}$ relation in
Eq. (1) as a SMPG plot named after Sturtevant, 
Murphy, Privalov, and Gill.$^{10,15}\ $  Baldwin's theorem states
that a linear SMPG plot is equivalent to having a convergence
temperature for $\Delta S_i$, and the intersection $a$ in the 
SMPG plot is the unique $\Delta S^*$.  Conversely, the existence of 
a convergence temperature indicates a linear SMPG plot at every 
temperature.

\vskip 0.3 cm
\SN {\it Lee's Theorem}

	Lee's theorem$^{13}\ $ is an application of the mathematical 
property of a bilinear function: if a function $f(x,y)$ is linear 
as a function of either $x$ or $y$, as well as a function of their
product, $xy$, then there exists a $y^*$ at which $f(x,y^*)$ is
independent of $x$.  To express this by equation:
\[      f(x,y) = a + bx + cy +dxy
      = d(x+\frac{c}{d})(y+\frac{b}{d}) + \frac{ad-bc}{d}       \] 
so when $y = y^* = -b/d$, $f(x,y^*)$ is independent of $x$. 

	The bilinear function has to be a linear function of either
variables when the other one is at a fixed value.  To apply this
result to the thermodynamics of dissolution, we identify the logarithmic 
temperature and a certain molecular parameter as the variables $x$ and
$y$, and the entropy (or enthalpy or Gibbs free energy) of dissolution
as the function $f(x,y)$.  Hence, at a given temperature, the entropy
is a linear function of the molecular parameter $X$.  Lee proposed 
the parameter to be a size measure of the 
solute molecule.  In fact, Lee's proposal of 
\begin{equation}
                    \Delta S_i(T) = a_s + b_s X_i             
\end{equation}
is a structural interpretation of the linear SMPG plot given in (1).
The subscript ``s'' here stands for entropy.
(There is a similar relation for enthahlpy, with its respective $a_h$ 
and $b_h$.$^{13}\ $)
It is well known that $\Delta C_p$'s are proportional to molecular
size of a hydrophobic solute.  It is also worth noting that a
linear relationship due to substituents of a series of functional
groups is widely observed in organic reactions.$^{17}\ $

	Therefore, the bilinear argument immediately leads to the 
existence of a convergence temperature from the linear entropy relation 
in Eq. (3).  Combining Lee's and Baldwin's theorems, it can be shown 
that the linear entropy relation is sufficient but not necessary for 
generating a convergence temperature.

\vskip 0.3 cm
\SN {\it The BMDW Theorem} 

	This theorem, which was given by Baldwin and Muller,$^{18}\ $ and
also independently by Doig and Williams,$^{19}\ $ establishes an
intrinsic relationship between the three convergence temperatures for
entropy, enthalpy, and Gibbs free energy.  Consider a series 
of homologous reactions designated by $i$ and 
each has a temperature-independent $\Delta C_{p_i}$ of its own.  If
two out of the three thermodynamics quantities (entropy, enthalpy,
and Gibbs free energy) have convergence temperatures,
then there is a convergence temperature for the third quantity.
If we denote by $T^*_s$, $T^*_h$, and $T^*_g$ these convergence 
temperatures, we have:
\begin{equation}
               T^*_g - T^*_h -T^*_g ln(T^*_g/T^*_s) = 0
\end{equation}   
and also a relation for the corresponding thermodynamic quantities:
\begin{equation}
       \Delta G(T^*_g) = \Delta H(T^*_h) - T^*_g \Delta S(T^*_s) 
\end{equation}

	This theorem was used by Baldwin and Muller$^{18}\ $ to explain 
the intriguing fact of approximately equal convergence temperatures 
$T^*_s$ and $T^*_h$ for a set of protein folding reactions.  It is 
discovered that this set of proteins have approximately equal melting 
temperatures, i.e., $T^*_g$ = 331 $\pm$ 9 K and $\Delta G(T^*_g)$ 
$\approx$ 0.  Using the same theorem, Doig and Williams$^{19}\ $ have
reached a similar conclusion.  They pointed out that most proteins of
the same set have approximately equal $\Delta G$ per residue at room 
temperature.  These two arguments are consistent with each other
if we note that all the proteins in the set have approximately
same $\Delta C_p$ per residue.

\vskip 1.0cm \noindent
{\large {\bf A Physical Model Based on SPT}}
\vskip 0.5cm \noindent

	The basic results of SPT are summarized here.  It is not 
necessary for the readers to know the technical details of SPT;$^5\ $ 
rather, our model starts with Eqs. (6) and (7) below.  SPT is a rigorous 
statistical mechanical theory for liquids made of hard spheres.  The 
theory also provides an approximated formula for calculating the free 
energy of dissolving a hard spherical solute from gas phase into a hard 
spherical solvent, which is equivalent to introducing a spherical cavity 
in the hard sphere liquid.  The free energy, which is a function 
of the radius of solute as well as the density and the radius of pure
solvent, has three dominant terms.  The first term is independent of 
solute size; it is associated with the reduction of conformational space 
of the liquid upon introducing a solute molecule of zero physical size 
into the solvent.  With the physical point being present in the middle 
of the solution, no solvent molecule can occupy the same point in space.  
Hence such conformations are no longer accessible.  This term is a function 
of the density and the radius of pure solvent only.  The second term is 
proportional to the square of the radius of the solute, i.e., 
its surface area.  And the third term is proportional to the cubic 
power of the radius, i.e., its volume, and external pressure.  It 
has been repeatedly demonstrated, both theoretically and empirically,
that the volume term is negligible for any molecular size cavity, 
and the dominant effect is from the surface term which
defines a proportional coefficient called surface tension.  For
hard spherical solvent and solute, SPT gives an expression for the
surface tension, and also a minor correction term on the
free energy due to different curvature of the solute molecule.

	The free energy is given as:$^5\ $ 	
\begin{equation}
 \frac{\Delta G}{k_BT} =  -ln(1-\xi) 
   + \frac{r^2}{a^2} \left[ \frac{3 \xi}{1-\xi}
   + \frac{9}{2} \left( \frac{\xi}{1-\xi} \right)^2 \right] 
\end{equation}
where $k_B$ is the Boltzmann constant, $T$ is temperature in 
Kelvin, $a$ is the radius of solvent molecule, $\xi$ is the packing
density for pure solvent (the volume fraction occupied by the hard 
spheres), and $r$ is the radius of solute molecule. 

	Our model starts with a very simple premise:
\begin{equation}
        \Delta G(T) = \delta(T, \xi) + \sigma(T,\xi) r^2
\end{equation}                       
where $\delta$ and $\sigma$ are solvation energies for a point and 
surface tension, respectively. 
Thermodynamics for solvation will depend on the functional form
of $\delta$, $\sigma$, and implicit temperature dependence of $\xi$. 
Eq. (7) offers the possibility for connecting thermodynamics of 
solvation with the thermal expansion coefficient of pure solvent 
(see below).

	From Eq. (7), it is straightforward to obtain:
\begin{equation}
        \Delta S(T) = -\delta_T - \delta_{\xi} \xi_T
                    - (\sigma_T + \sigma_{\xi} \xi_T) r^2 
\end{equation}
\begin{equation}
   \Delta H(T) =  \delta - T \delta_T - T \delta_{\xi} \xi_T
              + ( \sigma - T \sigma_T - T \sigma_{\xi} \xi_T) r^2 
\end{equation}
where subscripts ``$T$'' and ``$\xi$'' represent partial derivatives
with respect to these variables. 

	Let's now estimate the magnitudes of the various terms using
values for water$^3\ $ (SPT parameters for water at room 
temperature are $\xi$ = 0.363 and $a$ = $r_w$ = 1.38 $\AA$.)  $\xi_T$ is 
proportional to the thermal expansion coefficient of the pure solvent: 
$\xi_T = -\alpha\xi$, where $\alpha$ for water is very small.  At 1 atm,
the values of $\alpha$ for water range from $-0.064 \times 10^{-3}$ at 
$0^{\circ}$C to $0.7 \times 10^{-3}$ at $100^{\circ}$C, and 
$\alpha$ = $0.257 \times 10^{-3}$ at room temperature ($25^{\circ}$C). 
The thermal expansion coefficients for non-associative liquids are around 
$1 \times 10^{-3}$ at 1 atm and between $0^{\circ}$C to $100^{\circ}$C, so 
these values are not very different from that of water.

	Quantitatively, compare Eqs. (6) and (7), we have:

\[   |\delta_{\xi} \xi_T| = \left| \frac{k_BT \alpha \xi}{1-\xi}
     \right| \approx k_B \left( 
    \frac{300 \times 0.257 \times 10^{-3} \times 0.363}{1-0.363} \right)
                          = 0.04 k_B                    \] 
while
\[    \delta_T = -k_B ln(1-\xi) = 0.5 k_B >> |\delta_{\xi} \xi_T|    \]
where we have used $T$ = $300 K$.  Similarly by simple differentiation,
\[    \sigma_{\xi} = k_BT \frac{3(1+2\xi)}{r_w^2(1-\xi)^3}
                   = 20 k_BT / r_w^2.                     \]
Hence $|\sigma_{\xi} \xi_T|$ $\approx$ $0.56k_B / r_w^2$, while $\sigma_T$
= $3.17 k_B / r_w^2$ $>>$ $|\sigma_{\xi} \xi_T|$.

\vskip 0.3 cm
\SN {\it The Heat Capacity $\Delta C_p$}

	The heat capacity can be obtained from either Eq. (8) or 
Eq. (9): 
\begin{equation}
    \Delta C_p = -T ( \delta_{TT} + 2 \delta_{T\xi} \xi_T
               + \delta_{\xi\xi} \xi_T^2 + \delta_{\xi} \xi_{TT} )
               - T ( \sigma_{TT} + 2 \sigma_{T\xi} \xi_T
               + \sigma_{\xi\xi} \xi_T^2 + \sigma_{\xi} \xi_{TT} ) r^2
\end{equation} 

	While the values for $\alpha$ are not very different between
water and non-associative liquids, there is a dramatic difference 
between $\alpha$'s dependence on temperature.  This difference
contributes a large term to the $\Delta C_p$ of solvation in water, in 
contrast to non-associative solvent.  In other words, even though 
$\alpha$ for water is very small, its temperature dependence is
quite large, in contrary to most organic solvents.  Hence we will 
neglect contribution from $\xi_T$ but shall keep the terms with 
$\xi_{TT}$.  Thus we have:
\begin{equation}
    \Delta C_p = -T \xi_{TT} ( \delta_{\xi} + \sigma_{\xi} r^2 )
\end{equation} 
where, according to SPT (i.e., Eqs (6) and (7)), both $\delta_{\xi}$ and 
$\sigma_{\xi}$ are explicitly proportional to T.  Therefore if $\xi_{TT}$ 
is proportional to $T^{-2}$, then $\Delta C_p$ will be approximately 
temperature independent. 

	It is clear from the above argument that, for a non-associative 
solvent, the solvation of a inert solute should have very small 
$\Delta C_p$ since non-associative liquid has almost zero $\xi_{TT}$.  
For water, $\delta_{\xi}$ = 1.57$k_BT$ and
$\sigma_{\xi} r^2$ = 20.0$k_BT (r/r_w)^2$.  Therefore, Eq. (11) gives 
a linear relationship between $\Delta C_p$ and molecular surface area 
$r^2$ with almost zero intersection when $r > r_w$.  Note, however,
that for hydrocarbon solutes like propane and isobutane, there 
are significant solute-solvent interactions which contribute to the 
overall $\Delta C_p$.$^6\ $  Therefore, our present result is only
semi-quantitative and has to be augmented with such interactions when 
applied to real experimental data.

\vskip 0.3 cm
\SN {\it $\Delta S^*$ and $\Delta H^*$}

	We now return to Eq. (8). According to Lee's theorem:$^{13}\ $
\begin{equation}
          \Delta S^* = -\delta_T - \delta_{\xi} \xi_T 
\end{equation} 
Note that in order to compare calculation from SPT with experimental
measurements, we have to obtain the entropy from experimental measurements
according to molarity concentration scale.$^{5,11}\ $
When this was done,$^{13}\ $ Lee found that the
calculation given by SPT compares favorably with experimental
results. 

	Similarly, we have: 
\begin{equation}
      \Delta H^* = \delta - T \delta_T - T \delta_{\xi} \xi_T
                 = - T \delta_{\xi} \xi_T  
\end{equation} 
the second equality is because $\delta$ is a linear function of
$T$ (see Eqs. (6) and (7)).  Numerically, $-T \delta_{\xi} \xi_T$ 
is about 0.034 $k_BT$, that is, 2.5 kJ/mol.  It should be noted
that Eq. (13) neglects contribution from solute-solvent interaction.
It is known that different hydrocarbons, for example aromatics
and aliphatics, have different $\Delta H^*$.  On the other hand, 
inert gases could be used as a test for the present model.

	Thus, according to our analysis, the converging values for entropy 
and enthalpy are the consequence of point solvation energy.  This is an 
interesting conjecture.  A rigorous statistical mechanical treatment 
of this problem seems possible, but has never been developed.  In 
general, point solvation energy is dependent upon whether it is solid, 
or liquid, or gas, from which the solutes are transferred.  A quantitative 
theory might be able to explain the small differences between the
three SMPG plots.$^{10}\ $  (Note: after correction according to
molarity scale, the intersection, $\Delta S^*$, for gas $\rightarrow$ 
water dissolution is increased by about 60 J per degree per mole, 
see ref. 10).     

\vskip 0.3 cm  
\SN {\it $T^*_s$ and $T^*_h$ }

	It is natural to suspect that the unique convergence temperature,
$T^*_s$, for entropy of all three different groups of dissolution
transferred from either solid, liquid, or gas, is due to some intrinsic
properties of water.  $T^*_s$ is the
temperature at which $\Delta S(T)$ in Eq. (8) equals 
$\Delta S^*$ given in Eq. (12).  That is:
\begin{equation}
                \sigma_T + \sigma_{\xi} \xi_T = 0
\end{equation}
i.e., 
\[      \xi_T(T^*_s) = - \alpha(T^*_s) \xi(T^*_s)
                     = - \sigma_T/\sigma_{\xi}.              \]
By an approximated calculation, $\sigma_T/\sigma_{\xi}$
= 0.158/T.  Hence:
\[        T^*_s \alpha(T^*_s) \xi(T^*_s) = 0.158.             \]
This is consistent with the laboratory measurements $T^*_s$ = 383 K,
$\xi$ = 0.363 and $\alpha$ = $1.1 \times 10^{-3}$. 

	Similarly for $T^*_h$, from Eqs. (9) and (13) we have:
\begin{equation}
          \sigma - T \sigma_T - T \sigma_{\xi} \xi_T  = 0.
\end{equation}
According to SPT, the first two terms cancel each
other.  The third term, as we have indicated, is indeed quite small.  
Thus unfortunately, Eq. (15) is buried in our various 
approximations and fails to provide an estimation for $T^*_h$.  A 
more accurate estimation is required to obtain the convergence 
temperature for $\Delta H$.

\vskip 1.0cm \noindent
{\large {\bf Discussion}}
\vskip 0.5cm \noindent

	With the physical insight provided by SPT, we now attempt
to answer some key questions concerning the thermodynamics of hydrophobic 
solvation and the hydrophobic effect.  

\vskip 0.3 cm \noindent
{\it 1) What is the Hydrophobic Effect?}

	Ever since Kauzmann's seminal paper on hydrophobic 
effect,$^{20}\ $ people have believed that hydrophobic effect is mostly 
due to the reorganization of hydrogen bonds among the solvent molecules
around the solute, and the contribution of direct interaction between 
solvent and solute is rather minimal.  So what is the relation between
the hydrophobic effect and the solvation thermodynamics for hydrophobic 
solute in water?  To address the question, one has to be precise about 
the meaning of ``hydrophobic effect''.  There is an experimental 
(thermodynamic) side
and there is a structural (theoretical) side of conventional wisdom on 
``hydrophobicity''.  The experimental side is that non-polar solutes 
in water have very low solubility and the dissolution has large 
heat-capacity changes, in contrast to the dissolution in organic solvent.  
The structural side is that hydrogen bonding arrangement has been 
altered when an non-polar solute is dissolved in water.    

	From our analysis, it seems that low solubility and 
large $\Delta C_p$ in fact stem from two distinct sources.$^{21}\ $ 
While the large $\Delta C_p$ 
is associated with rearrangement of water molecules, 
the low solubility is primarily due to the geometric properties of 
water molecules.  This suggestion is consistent with our understanding 
of entropy-enthalpy compensation, which says that the ability of 
rearrangement of solvent should only have minor effect on 
solubility.$^7\ $  The more dynamic aspect of water will 
be reflected only through quantities like entropy, enthalpy, and 
heat capacity.  The reorganization process within solvent contributes 
to entropy change through heat capacity $\Delta C_p$, which
in turn is related to the peculiar large temperature dependence
of $\alpha$, the thermal expansivity, of water.  Hence it seems 
legitimate to identify $\Delta C_p$ with the hydrophobic effect.

	However, could one simply identify the $\Delta C_p$ term as 
hydrophobic free energy?  This indeed is the central issue behind the 
work by Murphy et al.$^{10}\ $  It was, of course, well recognized 
that the $\Delta C_p$ term is not uniquely determined until an 
appropriate reference temperature(s) is chosen.$^{22}\ $ 
Murphy et al. proposed the using of $T^*_s$ and $T^*_h$ as reference 
temperatures,$^{10}\ $ and thereafter a $\Delta C_p$ term
was uniquely defined.  However, as we have seen, the existence
of convergence temperatures and their values are {\bf not} the 
hallmark of the reorganization of associative solvent.  We suspect  
that many other solvation processes might also have such convergent 
properties.  The basis for the existence of convergence temperatures
is Eq. (7), and it is clear that this equation is not unique for 
hydrophobic solvation (more discussion later).

\vskip 0.3 cm \noindent
{\it 2) What is the Role of Hydrogen Bond?}

	An inevitable objection to our approach from many
readers will be the complete neglect of hydrogen bond which has 
central importance in Kauzmann's structural model for 
hydrophobicity.$^{20}\ $  We would like to emphasize that we 
do accept the hydrogen bond in water as the {\bf structural 
base} for hydrophobicity, but we want to seek the specific 
{\bf thermodynamic aspect} or aspects of the hydrogen-bond 
structure which are responsible for the thermodynamics of hydrophobic 
solvation.  The large temperature dependence of $\alpha$ no doubt 
is a manifestation of hydrogen bonding reorganization in water, 
the ultimate source of hydrophobicity.  However, other aspects
of the water molecules might also be relevant or even crucial; for 
example, the tetrahedral chemistry of hydrogen bonding,$^{20}\ $ 
or more generally the non-isotropic pair-wise interaction between 
two water molecules,$^{23}\ $ and the high ratio between physical
volume and thermodynamic volume of water.$^3\ $  In our model, SPT 
indeed uses all these properties of water, though not explicitly.  
The fact is that water molecule in the SPT model has large
thermodynamic volume, i.e., low packing density, but at the
same time a small radius leads to high solvation number
around a cavity.$^3\ $  This indicates that the solvent
molecule is not isotropic, and there are preferences for these
molecules to surround a cavity, which is exactly the Kauzamnn's 
argument!  A crude analogy will be a wedge-shaped molecule, 
and that is quite consistent with water molecules.

\vskip 0.3cm \noindent
{\it 3) What is the basis of convergence temperature?}
As pointed out by Lee,$^{13}\ $ the presence of convergence 
temperature is due to some kind of linear free energy dependence on molecular 
substituents, i.e., Eq. (7). 

\vskip 0.3cm \noindent
{\it 4)  What determine the magnitudes of $\Delta S^*$ and $\Delta H^*$?}
They are determined primarily by the 
thermodynamics of solvation of zero size point solutes.

\vskip 0.3cm \noindent
{\it 5)  Why is there a large $\Delta C_p$?}
The $\Delta C_p$ of hydrophobic solvation stems from the peculiar large 
temperature dependence of thermal expansion coefficient,
$\partial \alpha /\partial T$. 

\vskip 0.3 cm \noindent
{\it 6) How to Obtain Molecular Interaction Energy From the Thermodynamic
Data?}

	This is an age-old question.  Twenty-five years ago, T.H. Benzinger 
proposed a new definition for enthalpy of chemical reaction.$^{24}\ $  His
argument was that for a chemical reaction with non-zero $\Delta C_p$, 
there would be no unique heat of formation for the reaction.  Since the 
true heat of formation is mechanical (athermal), Benzinger suggested to 
use $\Delta H(0)$ at zero Kelvin as the ``true'' heat of formation, and 
argued that the remains of free energy should be lumped into one term:
\[      \Delta W(T) = \Delta H(0) - \Delta G(T).      \]
When taking into account the fact $\Delta S(0)$ = 0, he obtained:
\[   \Delta W(T) = \int_0^T \Delta C_p(X) 
                            \left(\frac{T}{X} - 1\right) dX  \]
In some sense, what Benzinger did was similar to what
Murphy et al. did.  They were both trying to divide the total free energy 
into a part with direct (mechanical) interaction and the rest part 
with surrounding effect due to thermodynamics.  They both realized that 
$\Delta C_p$ term was related to the latter since it characterizes the 
fluctuation in enthalpy due to thermal agitation.  The crucial question 
is of course whether it is possible to find an appropriate reference 
temperature(s) based on purely thermodynamic analysis without any 
molecular model.  Chan and Dill recently have extensively discussed 
this issue, and they concluded that without a molecular model, purely
thermodynamic analysis would not provide much meaningful 
result.$^{12}\ $

\vskip 0.3 cm \noindent
{\it 7. The Validity of Using SPT to Model Hydrophobic Solvation.}

	Let's now reiterate the rationales for using SPT to model 
solvation in associative solvents. 
In SPT, the reorganization of solvent is considered implicitly 
through experimental data on $\xi$ as function of temperature.
This approach is consistent with the assertion that the ultimate
reason for reorganization in a solvent is its temperature dependence 
as a pure liquid.$^7\ $  The presence of hydrogen bonds between 
solvent molecules is manifested in the experimental data on water 
density and its temperature dependence.  SPT, of course, completely 
neglects the soft interaction between solute and solvent. 

	Our second defense is based on a recent analysis of how 
thermodynamic systems respond to small perturbations.$^{7}\ $ It has
been shown that if we classify thermodynamic quantities by the 
orders of derivative of free energy, there is a relationship between
thermodynamic of perturbation and the thermodynamics of unperturbed 
system.  Since one can treat solvation as a perturbation, 
one only needs the thermodynamics of one order higher for pure solvent 
in order to calculate the thermodynamics of solvation.  For example, 
$\alpha_T$ of solvent gives $\Delta C_p$ of solvation.  All the 
structural changes in hydrogen bonding will be captured in these 
thermodynamic quantity of pure water, and our analysis made 
use of them.

	Finally, we would like to emphasize that we are not
attempting to use SPT to model the properties of water; rather,
we are merely using SPT to relate the thermodynamics of
dissolution of an inert solute in water to that of pure water.  
In the past, many models for water which are based on multi-state 
of water conformation have been successful in providing calculations
for thermal expansion coefficient, $\alpha$, but have failed to
deal with its temperature dependence.$^{25}\ $  This 
situation is completely in accord with our analysis.

\vskip 1.0cm \noindent
{\large {\bf Acknowledgments}}
\vskip 0.5cm \noindent
	 
	I thank Buzz Baldwin, B.K. Lee, and John Schellman for 
many detailed and helpful discussion in the past several years.  I wish 
to specially dedicate this work to Professor Baldwin on the occasion of
his 70th birthday.  His pioneer work on helical peptide, protein
hydrogen exchange, DNA flexibility, and hydrophobic effect have been 
guiding lights for my research.

\vskip 1.0cm \noindent
{\large {\bf References}}
\def \SN{\vskip 0.2 true cm \noindent}
\vskip 0.5cm \noindent
1.  Stillinger, F. H. (1973) J. Solution Chem., 2, 141-158.  \SN  
2.  Postma, J.P.M., Berendsen, H.J.C., \& Haak, J.R. (1982) Faraday
Symp. Chem. Soc., 17, 55-67.  \SN 
3.  Lee, B. (1985) Biopolymers, 24, 813-825.  \SN 
4.  Morton, A., Baase, W.A., \& Matthews, B.W. (1995) Biochemistry, 
34, 8564-8575.  \SN 
5.  Reiss, H. (1965) Adv. Chem. Phys. 9, 1-84.  \SN 
6.  Lee, B. (1991) Biopolymers, 31, 993-1008. \SN 
7.  Qian, H. \& Hopfield, J.J. (1996) J. Chem. Phys., 105, 9292-9298. \SN 
8.  Grunwald, E. (1984) J. Am. Chem. Soc., 196, 5414-5420.  \SN 
9.  Ben-Naim, A. (1975) Biopolymers, 14, 1337-1355.  \SN 
10.  Murphy, K., Privalov, P.L., \& Gill, S.L. (1990) Science, 
247, 559-561.  \SN 
11. Ben-Naim, A. (1978) J. Phys. Chem., 82, 792-803.  \SN 
12. Chan, H.S. \& Dill, K.A. (1997) Ann. Rev. Biophys. Biomol.
Struct., 26, 423-457.  \SN 
13. Lee, B. (1991) Proc. Natl. Acad. Sci. USA, 88, 5154-5158.  \SN 
14. Privalov, P. (1979) Adv. Protein Chem. 33, 167-241.  \SN 
15. Sturtevant, J.M. (1977) Proc. Natl. Acad. Sci. USA, 74, 2236-2240.  \SN
16. Baldwin, R.L. (1986) Proc. Natl. Acad. Sci. USA, 83, 8069-8072.  \SN 
17. Leffler, L. \& Grunwald, E. (1963) {\it Rates and Equilibria of
Organic Reactions}, John-Wiley \& Sons, New York.  \SN
18. Baldwin, R.L. \& Muller, N. (1992) Proc. Natl. Acad. Sci. USA,
89, 7110-7113.  \SN 
19. Doig, A.J. \& Williams, D.H. (1992) Biochemistry, 31, 9371-9375.  \SN 
20. Kauzmann, W. (1959) Adv. Protein Chem., 14, 1-63. \SN 
21. Lee, B. (1994) Biophys. Chem., 51, 271-278.  \SN  
22. Becktel, W.J. \& Schellman, J.A. (1987) Biopolymers, 26,
1859-1877.  \SN 
23. Ben-Naim, A. (1992) {\it Statistical Thermodynamics for Chemists
and Biochemists}, Plenum Press, New York.  \SN 
24. Benzinger, T.H. (1971) Nature, 229, 100-102.  \SN 
25. Kauzmann, W. (1975) Colloq. Inter. C.N.R.S., 246, 63-71.  \SN

\vskip 1.0cm \noindent
{\large {\bf Appendix: A possible $\Delta H^*$ in gas to
water dissolution?}} 
\vskip 0.5cm \noindent

	Let's proceed with the dissolution of a gas into water
(g $\rightarrow$ w) by first liquefying the gas
(g $\rightarrow$ l) and then transferring the liquid into water
(l $\rightarrow$ w):
\[   \Delta H_{g \rightarrow w} = \Delta H_{g \rightarrow l}
                  + \Delta H_{l \rightarrow w}.                \]
There seems to be a convergence temperature for dissolution
enthalpy change from liquid $\rightarrow$ water at about 
2$0^{\circ}$C, and corresponding $\Delta H^*$ is about few 
KJ/mol.$^{16}\ $  According to Trouton's rule, we can write
\[   \Delta H_{g \rightarrow l} = -88 T_b \hspace{0.2cm} J/mol \]
It is well known that the boiling temperature $T_b$ is scaled with
molecular size, hence $\Delta H_{g \rightarrow w}$ and
$\Delta H_{l \rightarrow g}$ should have similar converging 
$\Delta H^*$ but at different temperatures.  Mathematically, if
we have:
\[    \Delta H_{l \rightarrow w} = \Delta H^* + b_1(T) r^2      \]
where $b_1$ = 0 when T = 2$0^{\circ}$C, and
\[      \Delta H_{g \rightarrow l} = b_2(T) r^2         \]
then:
\[  \Delta H_{g \rightarrow w} = \Delta H^* + [b_1(T)+b_2(T)] r^2   \]
the convergence temperature is the value of T at which $b_1 + b_2$ = 0.

\end{document}